\documentclass{elsart}

\usepackage{amssymb}

\begin{document}

\begin{frontmatter}



\title{Perspectives in Fundamental Physics in Space}


\author{Orfeu Bertolami}
\address{Instituto Superior T\'{e}cnico (IST), Departamento de F\'{i}sica, 1049-001 Lisbon, e-mail:
orfeu@cosmos.ist.utl.pt}

\author{Clovis Jacinto de Matos}
\address{ESA-HQ, EUI-AC, F-75015 Paris, e-mail: Clovis.de.Matos@esa.int}

\author{Jean Christophe Grenouilleau}
\address{ESA-ESTEC, HME-EOI, NL-2201 Noordwijk, e-mail: Jean-Christophe.Grenouilleau@esa.int}

\author{Olivier Minster}
\address{ESA-ESTEC, HME-GAP, NL-2201 Noordwijk, e-mail: Olivier.Minster@esa.int}

\author{Sergio Volonte}
\address{ESA-HQ, SCI-CA, F-75015 Paris, e-mail: Sergio.Volonte@esa.int}

\begin{abstract}
% Text of abstract
We discuss the fundamental principles underlying
the current physical theories and the prospects of further improving their knowledge through
experiments in space.

\end{abstract}

\end{frontmatter}

% main text
\section{Introduction}
\label{} General Relativity (GR) and Quantum Mechanics (QM) are
the most fundamental and encompassing physical theories of the
$XX^{th}$ century. They are the cornerstones of all developments
aiming to unify the four fundamental interactions of Nature,
strong nuclear, electromagnetic, weak nuclear and gravitational
forces; and to harmonize gravity with the quantum picture of the
world. GR explains the behaviour of space-time and matter on
cosmologically large scales and of very dense compact
astrophysical objects. It is the most accurate theory so far of
the gravitational interaction. QM on the other hand, accounts for
the behaviour of matter primarily at small scales ({\AA} and
below), and ultimately leads, together with Special Relativity, to
the so-called Standard Model of strong and electroweak
interactions that accounts for all the observable known forms of
matter. QM also describes macroscopic quantum phenomena like
superconductivity, superfluidity and Bose-Einstein condensation.
Despite the great success of these theories, finding ways to unify
them into a single framework is the only way to understand the
high-energy behaviour of gravity and to avoid that gravity is not
consistent with fundamental principles such as, for instance,
Heisenberg's Uncertainty Principle.

Attempts to unify in a single theory the four fundamental
interactions of nature, and to harmoniously merge GR and QM, have
led to a rich lore of new physical models such as Kaluza-Klein
theories, Supergravity, and to the most fecund String/M-theory
\cite{[Green-Schwarz-Witten]} whose complex implications are still
largely untested. Moreover, the conceptual differences between GR
and QM seem to require deep changes in the underlying assumptions
about the nature of the Universe. String theory suggests for
instance, that the basic building blocks of the Universe are not
point-like particles, but instead strings and membranes. It also
implies that space-time has a non-commutative character.
Differences between GR and QM can be better appreciated through
still unresolved issues such as:

\begin{itemize}
\item The spatial non-separability of physical systems due to the
entanglement of states in QM, versus the complete spatial
separability of physical systems in GR.

\item The Equivalence
Principle of GR, versus the Uncertainty Principle in QM which may imply in violations
of the Equivalence Principle, as for instance, in some string theory models.

\item The
possible non-unitary evolution of pure states into mixed states
due to the existence of black holes solutions in GR. That is, the
presence of black-holes might blur the evolution of observable
quantities that in QM is performed by unitary operators.
\end{itemize}

We should also remember that to a great extent, the enormous
technological progress achieved since the beginning of last
century in telecommunications, electrical engineering,
electronics, photonics, information technology, nuclear
technology, etc, stems from the deep understanding of the
electromagnetic and nuclear interactions at quantum level.
Similarly, it is logical to expect that any gravity-related
technology must rely on a comparable level of understanding of the
gravitational interaction from the quantum mechanical point of
view.

Clearly, as in any branch of physics, progress is achieved through
the interaction between theory and experiments, and for what
concerns GR and QM in particular, further experimental testing of
the theoretical predictions and foundations of these theories may
reveal the important insights necessary to reach a higher level of
conceptual knowledge.  This paper, argues that space missions may
play an important role in the quest for a unification theory and a
quantum theory of gravity when ground experiments are not
feasible. In this respect, it is interesting to mention the
example of cosmology. Driven by important developments in
theoretical thinking and a great amount of data gathered by
dedicated space observatories, observational cosmology has become
a blooming subject. Upgraded versions of the COBE mission
\cite{[COBE]}, such as WMAP \cite{[WMAP]} of NASA and in the
future Planck mission \cite{[Planck]} of ESA, have given or will
give origin to a burst of activity on the physics of the Cosmic
Microwave Background Radiation. Similarly the Compton Gamma Ray
Observatory \cite{[COMPTON]} of NASA, together with the INTEGRAL
(INTErnational Gamma Ray Astrophysics Laboratory) mission from ESA
\cite{[INTEGRAL]} have prompted new developments of gamma ray
astronomy, as have the various X-ray telescopes XMM-Newton
\cite{[XMM-Newton]}, Einstein \cite{[Einstein]}, ROSAT
\cite{[ROSAT]}, Chandra \cite{[Chandra]}, etc.) and above all, the
Hubble Space Telescope \cite{[Hubble]}, which has dramatically
widened up our view of the Universe. The next generation of space
telescopes and observatories will not cease to surprise us and
will continue to be our major sources of data and inspiration for
new and revolutionary ideas. It is fairly reasonable to assume
that fundamental physics in space will follow the same pattern. Of
course, developments of fundamental physics in space are
intimately connected with the areas of particle physics, and
experimental gravity, in particular through the search for
deviations from Newton's law on small scales (below $1 \ mm$).

\section{Testing Well Known Theoretical Predictions and
Foundations}

GR is based on the generalization of the Principle of Relativity,
assuming that the laws of Nature are independent of the state of
motion (uniform or accelerated) of the reference frame with
respect to which they are formulated. This principle provides the
foundation of the universality of the laws of physics as it
ensures that these are independent of the state of motion, and of
the space-time location of observers. This endows a democratic
status for all observers. The set of experiments sustaining the
generalized Principle of Relativity are the following:

1. Physical laws are independent of the position and velocity of
the frame of reference thanks to the invariance of the world-line
distance between events in the spacetime continuum. This is
ultimately related with the fact that the speed of propagation of
the electromagnetic and gravitational interactions is constant and
independent of the frame of reference. This speed is the speed of
light in the vacuum.

2.  The acceleration of a test body falling under the single
influence of the gravitational interaction is independent of its
mass. This can be understood only if inertial and gravitational
masses are exactly equal to each other.

The first set of experiments is associated with the invariance of
the physical systems under translations and rotations in spacetime
usually referred to as:
\begin{itemize}

\item
Local Lorentz Invariance (LLI) (independence of the frame of
reference velocity).

\item
Local Position Invariance (LPI) (independence of the position
of the frame of reference)

\end{itemize}

The second set of experiments concerns the so-called Weak
Equivalence Principle (WEP)or Strong Equivalence Principle (SEP)
when gravitational self interaction is important (see e.g.
\cite{[berto]}).

The Principle of Special Relativity establishes only the
equivalence between inertial reference frames relying on a global
version of the first set of experiments above. Therefore, it does
not encompass the gravitational interaction. A generalization of
the Principle of Special Relativity to include gravity allows for
a covariant formulation of this interaction.

The covariant formulation of gravity implies a set of dynamical
equations for the spacetime metric, the so-called Einstein field
equations. These equations express the geometric nature of the
gravitational interaction, and describe how matter/energy and
spacetime geometry influence each other.

In the limit of weak gravitational fields and low velocities
compared with the speed of light, GR yields small corrections to
Newtonian gravity through the addition of terms proportional to
$GM/rc^2$, where $G$ is Newton's gravitational constant, $M$ the
mass and $r$ the radius of the source of the gravitational field
under consideration. Thus, general relativistic corrections will
become important in the case of compact astronomical objects,
such as neutron stars ($GM/rc^2=O(10^{-1})$) and black holes
($GM/rc^2=O(1)$), and for the Universe as a whole.

\subsection{Detection of Gravitational Waves}

Einstein's field equations predict the existence of gravitational
waves, which correspond to quadrupole oscillations of the
spacetime continuum itself. Those have already been indirectly
detected in binary pulsar systems \cite{[Hulse-Taylor]} via
tracking of the Post-Keplerian Parameters of the system, and
comparison with GR.

LISA (Laser Interferometer Space Antenna) \cite{[LISA]} is a
particularly eloquent example of a space mission devoted to test
fundamental physical principles, through the detection of
gravitational waves. LISA consists of a swarm of three satellites
forming an equilateral triangle with sides of 5 million km. Each
satellite located at a vertex of the triangle emits a laser beam
to the other two satellites, so as to form with the phase-locked returm beams
an interference pattern in the optical modules on-board each spacecraft. When a
gravitational wave crosses the triangle, the interference pattern
is shifted by an amount proportional to the intensity and the
frequency and polarization of the incoming gravitational wave. ESA's LISA
pathfinder mission (formerly called SMART-2) will play a crucial
role in developing and testing some of the technological requirements of a
mission as sophisticate as LISA.

LISA will lead to a fundamentally new window for observing the
Universe through observation of sources of gravitational waves.
Astronomy has so far mostly observed the sources of
electromagnetic radiation in the Universe. Gravitational astronomy
will allow scientists to achieve a deeper understanding of the
dynamics of the cosmos since gravitational waves couple very
weakly with matter and therefore suffer little scattering and
absorption on the way from the source to the observer.

\subsection{Detection of Gravitomagnetism}

GR also predicts, in the weak field limit and at first order
beyond Newtonian gravity, that for certain mass configurations (a
current like one), the metric can be decomposed into two vector
fields. The first one, usually referred to as gravitoelectric
field, corresponds to Newton's gravitational field. The second
corresponds to a "new" field, the so-called gravitomagnetic field.
These designations arise from the fact that in this approximation,
Einstein's field equations can be formulated in a way that
resembles Maxwell's equations for the electromagnetic field.
Clearly, direct experimental detection of gravitational waves and
the gravitomagnetic field produced by Earth's rotation, are
important tests of GR.

ESA Hyper (Hyper precision cold atom interferometry in space)
concept \cite{[HYPER]} and the NASA Gravity Probe-B mission
\cite{[Gravity Probe B]}, a Stanford University mission which has
been launched last 20th April 2004, are dedicated to the detection
of Earth's gravitomagnetic field. The Gravity Probe B satellite
circles the Earth in a polar orbit at an altitude of 650 km. Data
taking was concluded in August 2005 and results are expected in
2007. The mission concept consists in using four spinning
gyroscopes and a telescope. The telescope has been pointed to a
guiding star, IM Pegasi, and the gyroscopes were electrically
induced to align parallel to the telescope axis. Over a year of
operation about 5000 orbits are expected. The gyroscopes are left
undisturbed as the telescope is kept pointing toward the guiding
star through attitude control thrusters of the spacecraft.
According to GR, the drift angle between the gyroscopes and the
telescope is about $6.6''$, due to the Earth's geodetic effect,
while a smaller angle of $0.041''$ should open up in the direction
of the Earth's rotation, due to the Lens-Thirring effect.

Hyper on its hand, aims to perform the measurement of the
gravitomagnetic field through the phase shift it causes in an
interferometry experiment involving cold atoms rather than through the
motion of macroscopic bodies. It is relevant to mention that Hyper is just a mission concept
and that most likely it will be made concrete through a mission such as ``Fundamental
Physics Explorer'' described at ESA's Cosmic Vision 2015-2025 \cite{[cosmic-vision]}, that
will provide the drag-free platform on board of which experiments of
interference of atomic beams and with Bose-Einstein condensates will be performed.

Another interesting mission concept to detect gravitomagnetism
involves the so-called gravitomagnetic clock effect. This effect
is based on the time difference caused by the gravitomagnetic
field between two high precision clocks orbiting the Earth in
clockwise and anti-clockwise directions \cite{[Mashhoon]}.

\subsection{Testing of Basic Assumptions of General Relativity}

Testing the basic conceptual assumptions of GR, represents an
important challenge for space fundamental physics. This involves
experimentally testing the WEP, LLI and LPI.

The WEP establishes a composition-independent limit on the free
fall of bodies. This means that in a gravitational field the
gravitational mass, cancels out with the inertial mass given their
equality. This equality is established with great accuracy and has
been tested since Galileo in $1590$, Newton in $1686$, Bessel in
$1832$ and so on until the current most stringent limit
\cite{[Adelberger]}:

\begin{equation}
\vert\frac{m_i-m_g}{m_i}\vert <5\times10^{-13}
\end{equation}

Ground based experiments designed to verify the WEP are limited by
the unavoidable micro seismic activity of Earth. Space experiments
offer the possibility of improving the precision of current tests
by a factor of $10^2$ to $10^5$.

MICROSCOPE (MICROSatellite \`{a} train\'{e} Compens\'{e}e pour
l'Observation du Principe d'Equivalence) \cite{[Microscope]} is a
collaborative CNES - ESA mission to be launched in 2009, designed
to evaluate the WEP through the monitoring of the free fall of two
pairs of masses orbiting the Earth located in a drag free
environment at room temperature. The measured signal is the force
required to keep the test masses in a pair centered on each other.
Microscope will evaluate the WEP with a precision expected to
reach 1 part in $10^{15}$.

Unfortunately, the more ambitious ESA/NASA STEP (Satellite Test of
the Equivalence Principle) \cite{[STEP]} mission is currently
being studied by NASA only. The drag-free STEP spacecraft was to
carry four pairs of test masses accommodated in a superfluid
He-dewar at $2 K$. Differential displacements between the test
masses of a pair would be measured by SQUID sensors, and the
expected precision with which the WEP would be tested was 1 part
in $10^{18}$. Such a level of precision would allow checking
constraints introduced by existing string theories
\cite{[Damour]}.

Another promising possibility for testing the WEP uses cold atom
interferometry. Ground based High-precision gravimetric
measurements have been made using the interferometry of
free-falling Cesium atoms, and allowed to reach a precision of 7
parts in $10^9$ \cite{[Peters]}. Ultimate precision of this method
can only be achieved in space. As an example, the resolution
provided by the atom interferometers to be used in ESA's Hyper
concept mission which could be sufficient to perform a test of the
WEP with an improved precision by a factor of $10^6$. Hyper would
carry two cold-atom Sagnac interferometers (based on the negative
Michelson-Morley experiment for detection of the ether drift). By
comparing the rate of fall of Cesium and Rubidium atoms in two
independent interferometers a precision of the order of $1$ part
in $10^{15}$ could be achieved, and this would represent an
independent confirmation of, or perhaps a disagreement with, the
results of MICROSCOPE. As already mentioned, a concrete mission to
fulfil Hyper concept will be the ``Fundamental Physics Explorer''
described at ESA's Cosmic Vision 2015-2025 \cite{[cosmic-vision]}.

Invariance under Lorentz transformations (LLI), which states that
the laws of physics are independent of the frame velocity, is one
of the most fundamental symmetries of physics and a basic
ingredient of all known physical theories. However, recently some
evidence has been found, in the context of String/M-Theory, that
this symmetry can be spontaneously broken. Naturally, this poses
the challenge of verifying this possibility experimentally. The
most accurate laboratory tests of LLI are performed via the
so-called Hughes-Drever experiment \cite{[Hughes]}
\cite{[Drever]}. In this type of experiment, one searches whether
there exists any anisotropy of inertia through the study of
resonant absorption of photons by a $Li^7$ nucleus in a strong
magnetic field. The ground state has spin $3/2$ and splits into
$4$ equally spaced energy levels, given that nuclear physics laws
are rotationally invariant. Therefore, if inertia is not
isotropic, then the four states will not remain exactly equally
spaced over the 12 hours period of Earth's rotation in which the
magnetic field is carried to two different locations with respect
to the galactic center. This technique allows achieving impressive
limits, the most stringent being \cite{[Lamoreaux]}.

\begin{equation}
\delta\equiv\mid\frac{m_I c^2}{\sum_{A} E_A}-1\mid <
3\times10^{-22},
\end{equation}

where $E_A$ are the relevant binding energies. From astrophysical
observations, limits on the violation of momentum conservation and
the existence of a preferred reference frame can be set from
bounds on the post-Newtonian parameter, $\alpha_3$ which vanishes
identically in GR. It can be determined from the pulse period of
millisecond pulsars \cite{[Will]}, \cite{[Bell]}. The most recent
limit, $\alpha_3<2.2\times10^{-20}$ \cite{[BellD]}, indicates that
Lorentz symmetry holds up to this level. We should mention that,
in broad terms, in the Parametrized Post-Newtonian Formalism the
metric is expanded in powers of the Newtonian potential, velocity
of matter and velocity with respect to a preferred frame. Clearly,
the presence of the latter implies in the breaking of translation
and/or rotational invariance and hence, yielding that momentum
conservation is violated.

It is known that the propagation of the ultra-high-energy protons
is limited by inelastic collisions with photons of the Cosmic
Microwave Background radiation making it impossible to protons
with energies above $5\times10^{19} eV$ to reach Earth from
distances farther than $50-100 \ Mpc$. This is the so-called
Greisen-Zatsepin-Kuzmin (GZK) cutoff \cite{[Greisen]}. However,
events where the estimated energy of the cosmic primaries is
beyond the GZK cutoff have been observed by different
collaborations \cite{[Yoshida]} \cite{[Bird]} \cite{[Brooke]}
\cite{[Efimov]}. The issue is controversial. For instance, for the
AGASA collaboration \cite{[AGASA]} this is only a $2.2\sigma$
effect. The confirmation of these observations by the most recent
HiRes collaboration is still under debate \cite{[Abbasi]}. Despite
that, it has been suggested \cite{[Coleman]} that slight
violations of Lorentz invariance would cause energy-dependent
effects which would suppress otherwise dynamically inevitable
processes, e.g. the resonant scattering reaction,
$p+\gamma_{2.73K}\longrightarrow\Delta_{1232}$, where
$\Delta_{1232}$ is the $1232~MeV$ hadronic resonance. The study of
the kinematics of this process allows to set quite stringent
bounds on the degree to which Lorentz invariance holds,
$\delta_{Lorentz}<1.7\times10^{-25}$ \cite{[Coleman]}
\cite{[Bertolami1]} \cite{[Bertolami2]}.

In what concerns LPI, experiments on the universality of the
gravitational red-shift set the measure to which this symmetry
holds. Hence, violations of the LPI would imply that the rate of a
free falling clock would be different when compared with a
standard one, for instance on the Earth's surface. Thus, one of
the most accurate determinations of the LPI has been achieved from
the comparison of hydrogen-maser frequencies on Earth and on a
rocket at $10000 \ km$ altitude \cite{[Vessot]}. The most recent
band is about $2\times10^{-5}$ of the Newtonian potential divided
by the velocity of light square.

On very large scales, the Hot Big-Bang Model describes the
Universe through GR and the assumption that matter and radiation
are homogeneously and isotropically distributed. Compatibility
with data suggests that we are living in an accelerating, low
matter-density Universe. The origin of this acceleration can be
due to either to a cosmological constant \cite{[Bento1]}, or a
slow-varying vacuum energy of some scalar field, usually referred
to as Quintessence \cite{[Caldwell]}, or due to an exotic new
equation of state, the generalized Chaplygin equation of state
\cite{[Bento2]}. This dark energy amounts for a substantial part
of the energy density of the Universe, $\Omega_\Lambda\simeq0.73$,
with the contribution from matter, dark\footnote{Most likely
candidates for dark matter include a linear combination of neutral
supersymmetric particles, the neutralinos (see eg.
\cite{[Bottino]}),  axions \cite{[Asztalos]} and a
self-interacting scalar particle \cite{[Bento3]}.} and baryonic,
$\Omega_{DM}\simeq0.23$, $\Omega_{Baryons}\simeq0.04$, so that
$\sum_{i} \Omega_i=1$ but with no contribution from the spatial
curvature \cite{[Perlmutter]} \cite{[Garnavich]} \cite {[b3]}
\cite{[Spergel]} as predicted from Inflation (see for instance,
\cite{[Olive]}).

Thus, at late times the rate of expansion of the Universe is
controlled by the dark energy component, which has negative
pressure. It should be mentioned that the understanding of the
quantum properties of vacuum and how it relates with the observed
value of energy density are amongst the greatest challenges for
$XXI^{st}$ century physics.

\subsection{Testing quantum Mechanics in Space}

Space platforms, such as the "Fundamental Physics Explorer" that
is part of ESA's long-term scientific objectives, a "cosmic
vision" \cite{[cosmic-vision]}, also offer a unique drag-free
environment to investigate the predictions of quantum physics. A
test of quantum entanglement over astronomical distances would
indeed be of great scientific value. The evaluation of the
influence of gravity on quantum entanglement, and therefore the
possible use of quantum entanglement to investigate the quantum
features of gravity at low energies \cite{[Tajmar1]} are issues
that deserve more investigation. Space experiments involving
entangled systems over large distances and different gravitational
environments, are particularly well suited to convey this type of
research.

On the other hand, experiments such as EUSO \cite{[EUSO]} and
LOBSTER onboard the ISS for the space observation of cosmic rays
with energies greater than the ones achievable in particle
accelerators, will also help to push even further our
understanding of high-energy physics. Notice that the EUSO
experiment has been postponed until ground-based cosmic-ray
observatories like AUGER \cite{[Auger]} yield results.

Testing QM in space is also very important for the future use of
novel technologies that will rely entirely on the unusual features
of the quantum world. Emerging fields like spintronics,
nanotechnology, quantum computing and quantum communication
\cite{[Zeilinger]} will certainly represent new technological
opportunities to expand the possibilities of spaceflight.
Nevertheless, these technologies that are still under development
on ground, will need proper qualification for possible use in
space. Therefore, quantum physics experiments in space will not
only provide deeper insights; through fundamental physics missions
we will also acquire experience needed to fulfill these
qualification steps in the future.

\section{Investigating Phenomena not Clearly Encompassed by
Theory}

Controversy sparked by theoretical thinking and consensus reached
through experimentation is the engine of science. A scientific
revolution is most often initiated when a new experimental result
does not properly fit within the accepted physical theories.
According to the science philosopher Thomas Kuhn, the emergence of
a new paradigm occurs due to the resistance of the scientific
community in accepting at first a new physical picture to explain
unexpected experimental results. What are the experimental
anomalies and/or theoretical issues in GR and QM that might lead
to new insights towards the goal of unifying in a single
theoretical frame, the fundamental interactions of Nature and
finding a suitable quantizing scheme for gravity? In what follows
we shall discuss two issues that we regard as being of particular
relevance.

1.  Celestial mechanics has been for centuries the main source of
discoveries in gravitational physics, from Kepler's laws to the
subtle anomalies of Mercury's orbit. Recently discovered anomalous
trajectories of the Pioneer 10 and 11, Ulysses and Galileo probes
seem to indicate that some anomalous gravitational-type force with
range beyond several $20$ AU or so might exist \cite{[Anderson]}.

2.  Analysis of the free fall of physical systems is, as already
discussed, a privileged experimental tool to test GR. It is
remarkable in this respect, that the free fall of electrically
charged particles and of antimatter has been so far poorly
investigated. It is extremely relevant that a novel round of free
fall experiments is carried out for charged particles and
antimatter.

Given the importance of these two issues we discuss them in more
detail next.

Of course, other anomalies related with the experimental
determination of an unexplained excess of mass of Cooper pairs in
superconductors\cite{[Tajmar2]} could be pointed out, however we
feel that their implications are not so clearly related with our
goal of discussing main fundamental physics questions that can be
studied in space.

\subsection{Testing the Weak Equivalence Principle for Antimatter}

The testing of the WEP for antiparticles remains still a largely
open problem, despite recent developments in producing an
appreciable number of antihydogen atoms by the ATHENA and ATRAP
collaborations at CERN \cite{[ATHENA]} \cite{[ATRAP]}. It is
somewhat urgent that free fall experiments for antimatter are
conducted so as to evaluate to which extent gravity complies with
CPT symmetry. This is a fundamental symmetry of quantum field
theory and corresponds to invariance of three conjugate
operations, where C stands for charge conjugation, P for parity,
and T for time reversal. In case gravity respects this symmetry,
antimatter will fall exactly like matter in a gravitational field.
From the experimental point of view, it should be mentioned that
special Penning trap devices, magnetic containers, were developed
for the purpose of storing substantial amounts of antimatter over
a long time. In this respect, experimental proposals like WEAX
(Weak Equivalence Antimatter experiment) \cite{[Lewis]} to be
conducted at a cryogenic vacuum facility onboard the ISS and which
aim to measure the free fall of antiprotons while orbiting the
Earth are particularly appealing. The main idea behind this type
of experiments is that antiprotons can be confined for a few weeks
in a Penning trap with a geometrical configuration in which the
effect of gravity would manifest itself as a perturbation on their
motion. The expected precision of the experiment is 1 part in
$10^6$, three orders of magnitude better than for a ground
experiment. Naturally, testing the gravitational properties of
antihydrogen as well as its spectroscopy, will allow a deeper
understanding of this symmetry. It is worth mentioning that in
some String-Field-Theory models, CPT symmetry can be spontaneously
broken, meaning that although it is a symmetry of the theory, it
is not shared by its ground state.

It is important to point out that these experiments have a high
scientific value as they can provide relevant insights on
extensions of the Standard Model. In this context, it is
interesting to remark, that free-fall experiments with charged
particles are also particularly relevant given the fact that they
are very poorly tested experimentally. In the case of ground-based
experiments, they involve, at least for the electron and the
positron, the Schiff-Barnhill effect \cite{[Schiff]} (see in Ref.
\cite{[Bertolami-Tajmar1]}, \cite{[Dittus]}).

\subsection{A Novel Intermediate Range Fundamental Interaction of
Nature?}

The investigation of the existence of new intermediate range
interactions of Nature at scales beyond $20$ AU, is another open
question that awaits a dedicated mission. An alternative to a
dedicated mission, would involve a somewhat more limited
experiment mounted as piggy-bag onboard deep space missions like,
for instance the Pluto-express mission or the NASA Interstellar
Probe mission \cite{[Interstellar Probe]}.

A putative new fundamental interaction was first considered by
Anderson and collaborators \cite{[Anderson]} in order to explain
the anomalies in the trajectories of the probes Pioneer 10 and 11,
Ulysses and Galileo, that imply the presence of an acceleration of
the order of $8\times10^{-10} m/s^2$ directed towards the Sun, and
that starts manifesting itself at distances beyond $20$ AU from
the Sun, after the influence of solar radiation becomes
negligible. This is the so-called Pioneer anomaly. This additional
interaction would manifest itself as being a different kind of
gravity with a coupling constant a fraction of Newton's
gravitational constant and a finite range. Its finite range
suggests the vector boson of this new interaction has a
non-vanishing mass that leads to a Yukawa-type term to be added to
the Newtonian gravitational potential. Thus, this subtle deviation
from Newtonian gravity could be attributed to the existence of a
new force of Nature. This force would in turn lead to violations
of the WEP through deviations of the universality of free fall.

It is debatable whether the Pioneer anomaly is due to some
un-modeled engineering problem of the probes, or whether it
signals new physics (see Refs. \cite{[Bertolami3]} and
\cite{[Bertolami-Paramos1],[Bertolami-Paramos2],[Bertolami-Paramos3]}
for a discussion on the theoretical side of the matter. An
engineering solution is discussed in Ref. \cite{[Scheffer]}). The
demonstration that the gravitational field of Kuiper Belt is not
the cause of the anomaly has been recently reanalysed (see
\cite{[Bertolami-Vieira]} and references therein). In any case, it
is only through a dedicated deep space tracking experiment that
this phenomenon will be more clearly characterised and the issue
definitely settled.

A dedicated mission would in its simplest form consist in
launching into deep space a spherical probe whose behaviour
(mechanical, thermal, electromagnetic, etc.) is very well known
\cite{[Bertolami-Tajmar1]} \cite{[Bertolami-Tajmar2]}. Accurate
tracking of its orbit would allow for precise evaluation of the
Pioneer anomaly, as any deviation from the predicted trajectory
would be used to evaluate the un-modeled Pioneer acceleration.
Alternative mission concepts were discussed in Refs
\cite{[Nieto-Turyshev]}, \cite{[Johann]}. The use of laser ranging
techniques and the flying formation concept to the test the
Pioneer anomaly were recently discussed
\cite{[Pioneer-collaboration]}.

\section{Conclusion}

For more than half a century, classical physics has provided the
knowledge required to propel and transport manned and unmanned
missions throughout the Solar System. Contemporary physics
however, has not so far played a similar role. Advances in quantum
and relativistic mechanics were not yet fully implemented so to
lead to propulsion breakthroughs and to allow for a more efficient
exploration and utilization of space.

It is not inconceivable that the crisis of contemporary physics
may be partly responsible for this state of affairs since our
pictures of the world on very large and on very small scales do
not quite fit together.

Suitable space platforms can provide the proper drag free
environment for carrying out research in many critical areas of
modern physics. It is an exciting prospect to think that
fundamental physics missions in space may provide important
insights into the nature of the theory still to emerge that would
harmoniously encompass QM and GR. In turn, unification and a
synthesis of QM and GR may lead to technological breakthroughs
that will further push the boundaries of current space systems.

It is often said that quantum gravity is the most challenging
synthesis to be achieved in $XXI^{st}$ century physics. Even
though, the technological spin-offs of that theory are not clearly
visible yet, it may most probably change our society as former
scientific revolutions did in the past. Securing the steps to
ensure such a paradigm shift, culturally and technologically, is
in our view, an inescapable issue.

\section{Acknowledgement}

\noindent
The authors would like to thank J. P\'aramos and
A. Rathke for useful suggestions on the topics of this paper.

% \appendix

% \section{}
% \label{}

\newpage
\appendix
\section{Nomenclature List}

\noindent
\\AGASA: Akeno Giant Air Shower Array (Japan)
\\ATHENA: AnTiHydrogEN Apparatus (CERN)
\\ATRAP: Antihydrogen Trap Collaboration (CERN)
\\Auger: Pierre Auger Cosmic Ray Observatory (Argentina)
\\AU: Astronomical Unit
\\CERN: Centre Européen de Recherches Nucléaires
\\CNES:Centre National D'\'{e}tudes Spatiales (France)
\\COBE: COsmic Background Explorer (NASA)
\\CPT: Charge-Parity-Time Reversal
\\ESA: European Space Agency
\\EUSO: Extreme Universe Space Observatory (ESA)
\\HiRes: High Resolution Fly's Eye Collaboration
\\HYPER: HYPER-precision cold atom interferometry in space (ESA)
\\INTEGRAL: INTErnational Gamma Ray Astrophysics Laboratory (ESA)
\\ISS: International Space Station
\\LISA: Laser Interferometer Space Antenna (ESA/NASA)
\\LLI: Local Lorentz Invariance
\\LOBSTER: All-Sky X-Ray monitor (ISS)
\\LPI: Local Position Invariance
\\MICROSCOPE: MICROSatellite \`{a} tra\^in\'{e} Compens\'{e}e pour
l'Observation du Principe d'Equivalence (CNES-ESA)
\\NASA: National Aeronautics and Space Administration (USA)
\\QM: Quantum Mechanics
\\ROSAT: Roentgen Satellitte (Germany, UK, USA)
\\STEP: Satellite Test of the Equivalence Principle (NASA)
\\SMART: Small Missions for Advanced Research in Technology (ESA)
\\GP-B: Gravity Probe-B (NASA)
\\GR: General Relativity
\\WEAX: Weak Equivalence Antimatter Experiment
\\WEP: Weak Equivalence Principle
\\WMAP: Wilkinson Microwave Anisotropy Probe (NASA)
\\XMM: X-ray Multimirror Mission (ESA/NASA)

\end{document}